\documentclass{elsart3p}

\usepackage{graphicx}

\usepackage{amssymb}

\begin{document}

\begin{frontmatter}



\title{Investigation of the tunneling spectra in HgBr$_{2}$-intercalated 
Bi-2212 single crystals below and above T$_{c}$}


\author[a]{C. Kurter\corauthref{cor}}
\corauth[cor]{Corresponding author. Tel.: +1-630-252-8457}
\ead{kurter@anl.gov}
\author[a]{D. Mazur}
\author[b]{L. Ozyuzer}
\author[a]{D.G. Hinks}
\author[a]{K.E. Gray}

\address[a]{Materials Science Division, Argonne National Laboratory, Argonne IL 60439 USA}
\address[b]{Department of Physics, Izmir Institute of Technology, TR-35430 Izmir Turkey}

\begin{abstract}
Interlayer tunneling spectroscopy measurements were performed on mesa arrays of Bi-2212 single 
crystals, intercalated with HgBr$_{2}$.  Tunneling conductances were obtained over a 
wide temperature range to examine the spectral features, especially the behavior 
of the quasiparticle peaks corresponding to superconducting energy gaps (SGs).  
Experimental spectra showed that gap-like features are still present even for the temperatures 
far above the transition temperature, T$_{c}$.  This evidence is consistent with the idea that the 
SG evolves into a pseudogap above T$_{c}$ for HgBr$_{2}$-intercalated Bi-2212 single crystals.

\end{abstract}

\end{frontmatter}


One of the puzzles for high-T$_{c}$ superconductors is the pseudogap (PG) that results in a depression of density of states 
at the Fermi level above T$_{c}$ \cite{alff}. Since tunneling spectroscopy can reveal the states near the Fermi level, it can give 
an insight about PG apart from superconducting gap (SG) in the quasiparticle excitation spectrum.  
The PG is clearly observed to exist in underdoped and optimally doped Bi-2212 single crystals.  
However, it is relatively hard to see these features above T$_{c}$  for heavily overdoped Bi-2212 specimens, 
since SG and PG are barely distinguishable for these crystals \cite{ozyuzer}. 
In this study, we report the tunneling measurements of overdoped, HgBr$_{2}$-intercalated Bi-2212 
crystals with bulk T$_{c}$ $\sim$ 74 K.  Although interlayer tunneling is a perfectly convenient technique for 
studying naturally occurring intrinsic Josephson junctions (IJJs) within copper oxide superconductors, 
excess Joule heating because of poor thermal conductivity of these materials distorts the measured 
results substantially, e.g., possibly leading to a reduction in magnitude of SG.  Intercalating inert 
HgBr$_{2}$  molecules between adjacent Bi-O layers increases the c-axis resistivity resulting in a hundred-fold 
suppression of the heat dissipation at a given voltage per junction \cite{yurgens}. 

Pristine Bi-2212 single crystals were grown by a floating zone technique and intercalation was accomplished 
by a vapor transport reaction between pure Bi-2212 crystals and HgBr$_{2}$ in air at 230 $^{o}$C for 16 hours.  
After intercalation, T$_{c}$ for these crystals was measured to be 74 K, and it is generally thought 
that intercalated crystals are overdoped \cite{yurgens99}.  After additional heat treatments during mesa 
fabrication (150 $^{o}$C for 1 hour in flowing air) these crystals may have further overdoped to an even lower 
T$_{c}$ (see below).  
For even greater minimization of heating effects, small mesa structures (10x10 $\mu m^{2}$ by $\sim$50 nm 
height) were 
patterned on the crystal surface using photolithography and Ar ion milling \cite{kurter}. For interlayer tunneling 
measurements, two conventional contacts were made on two ends of the crystal while the third contact 
used a mechanically sharpened gold wire of 100 $\mu$m in diameter, making a gentle contact on top of a 
particular mesa of the array.  In order to see clearly the hysteretic quasiparticle branches in I-V 
characteristics, the current was cycled back and forth many times.

\begin{figure}
\begin{center}
\includegraphics*[bb=102 317 488 646, width=7.4cm]{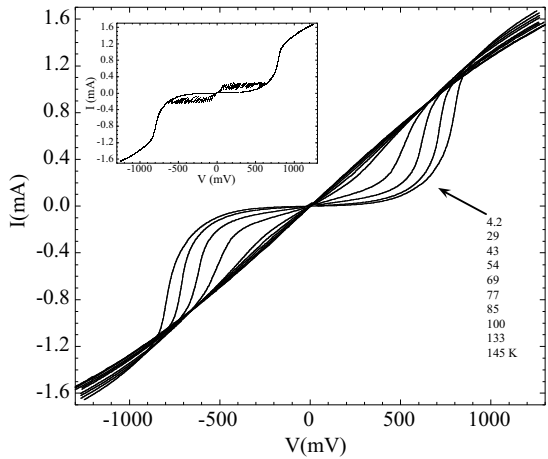}
\end{center}
\caption{Temperature dependence of I-V characteristics of a 10x10 $\mu m^{2}$ mesa. 
The inset shows I-V characteristics 
with quasiparticle branches at 4.2 K.}
\label{fig:exmp}
\end{figure}

Figure 1 shows the temperature evolution of the current-voltage characteristics: for clarity, multiple quasiparticle 
branches (see IJJ stack data obtained at 4.2 K in inset) were removed to emphasize the behavior of the 
sum-gap branches with increasing temperature.  We never observe ohmic behavior above T$_{c}$ (even at 145 K) 
out to the modest voltages measured.  The inset shows 30 individual tunneling junction resistive branches for 
this mesa with a spacing of $\sim$24 mV each.  

The tunneling conductances, i.e. the numerical derivatives of I-V curves shown in Fig. 1, are displayed in 
Figs. 2(a) and (b) at various temperatures from 4.2 K to 145 K. The quasiparticle peaks decrease in height and 
smear out with increasing temperature. However, the data for 77 to 112 K still show a broad gap-like quasiparticle 
peak structure that may be taken as evidence of a pseudogap in HgBr$_{2}$ doped Bi-2212 samples.  This feature can give 
an insight about the emergence of a pseudogap below 112 K.  Above 145 K the conductivity monotonically decreases 
with absolute voltage.  
The inset of Fig. 2(a) shows the gap voltage obtained from the peak position, V$_{p}$, vs. temperature from which 
a T$_{c}$ of $\sim$ 69 K 
can be inferred.  From the same plot one can judge the compatibility of the experimental data to the 
BCS temperature 
dependence.  Below T$_{c}$, the voltage drops much more rapidly with increasing temperature in the 
experimental data, and is never zero at T$_{c}$ . 
There is an upward deviation beyond T$_{c}$  that may be a sign of hybridization of superconducting 
peak with the pseudogap \cite{anagawa}. 
In conclusion, our interlayer tunneling measurements of HgBr$_{2}$  intercalated 
Bi-2212 single crystals endorse the existence of 
PG above T$_{c}$ $\sim$  69 K that we have observed up to 112 K.  

\begin{figure}
\begin{center}
\includegraphics*[bb=55 45 510 690, width=7.4cm]{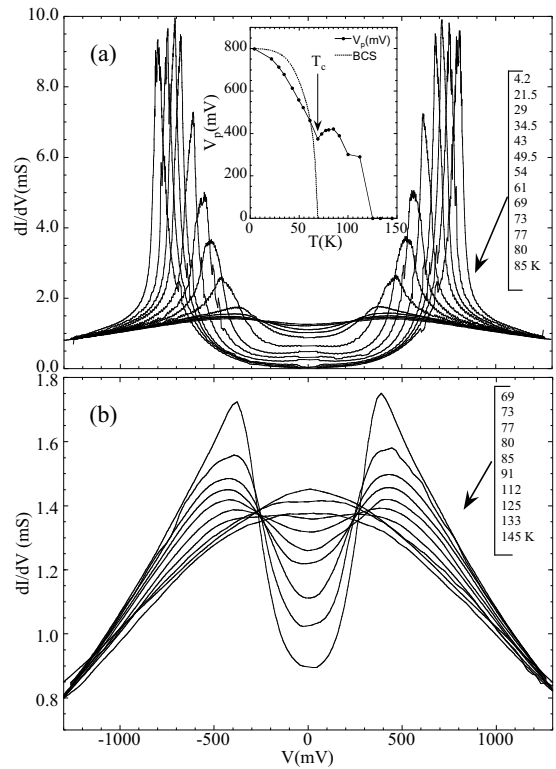}
\end{center}
\caption{Temperature dependence of dI/dV-V characteristics of a 10x10 $\mu m^{2}$ mesa. 
(a) T=4.2-85 K, (b) T=69-145 K.
The inset shows evolution of sumgap peaks with temperature.}
\label{fig:exmp}
\end{figure}

\end{document}